# Dual-label Deep LSTM Dereverberation for Speaker Verification

*Hao Zhang[1], Stephen A. Zahorian[1], Xiao Chen[1], Peter Guzewich[1], Xiaoyu Liu[2]*

[1]Electrical and Computer Engineering Department, Binghamton University, NY, USA
[2]Sony Interactive Entertainment Inc

{hzhang20,zahorian,xchen49,peter.guzewich}@binghamton.edu, xiaoyu.liu@sony.com

## Abstract

In this paper, we present a reverberation removal approach for speaker verification, utilizing dual-label deep neural networks (DNNs). The networks perform feature mapping between the spectral features of reverberant and clean speech. Long short term memory recurrent neural networks (LSTMs) are trained to map corrupted Mel filterbank (MFB) features to two sets of labels: i) the clean MFB features, and ii) either estimated pitch tracks or the fast Fourier transform (FFT) spectrogram of clean speech. The performance of reverberation removal is evaluated by equal error rates (EERs) of speaker verification experiments.

## 1. Introduction

Speaker verification is the task of determining whether a speaker's claimed identity is true by processing the speech audio. The accuracy of such a task, as well as automatic speech recognition (ASR), suffers when the audio is corrupted by reverberation, which occurs whenever the audio is obtained from a distant speaker [2][5]. Since reverberant conditions are common, dereverberation methods are of great interest.

One way to reduce degradation caused by reverberation is to map the reverberant speech representation to its clean counterpart, assuming reverberant speech and its corresponding clean version are both available for training. In [9], an algorithm named SPLICE was used for feature compensation which is essentially a linear mapping between reverberant and clean speech features. To train a nonlinear mapping or transform, neural networks (NNs) have long been utilized for speech enhancement [12]. With the advent of deep neural networks (DNNs) which are capable of highly comprehensive learning, nonlinear transforms of speech features improved considerably. An example of this approach for ASR is the state-of-the-art acoustic modelling used in [6]. A relatively recent variant of DNNs, deep long short term memory recurrent neural networks (LSTMs) have been reported to give better ASR accuracy than traditional DNNs for feature enhancement [13].

Deep LSTMs are exploited for dereverberation in the present paper and evaluated for the task of speaker verification. We used the bidirectional LSTMs (BLSTMs) structure, as used in [11][13], for its capacity to use both long-term past and future speech feature information to predict clean features for each point in time.

Inspired by the multi-task deep learning research for speech applications [2] where an additional training objective improves the effectiveness of the primary goal, the current study is based on dual-label BLSTMs. The idea is to have two sets of targets during training, so that weights of the network are trained for both sets of targets. During training, the inputs are reverberated Mel filterbank (MFB) outputs and the primary targets are clean MFB outputs, while either a clean pitch track, or clean fast Fourier transform (FFT) spectrogram, serve as a secondary target. [4] shows a spectrogram-to-MFB DNN mapping outperforms either a spectrogram-to-spectrogram or MFB-to-MFB mapping in terms of robust ASR. The mapping across different frequency domains is also probed in current work. In contrast to [4], the proposed method performs both MFB-to-spectrogram and MFB-to-MFB mappings simultaneously. The YAAPT pitch estimator is used to make the clean pitch targets [14].

## 2. Dual-label LSTMs

LSTMs were developed as an improved version of Recurrent Neural Networks (RNNs) in that the gradient vanishing problem is mitigated. In the current work, which uses deep LSTMs, a multi-label approach has been implemented and tested using the PyTorch Library [10]. The idea is to have two sets of targets during training, so that weights of the network are optimized for both sets of targets. Many recent studies demonstrate that a related second task can improve the training for the original task, and hence improve its performance [2].

In Figure 1, every blank rectangle block is an LSTM unit. The primary goal is to map the corrupted log scaled MFB outputs to underlying clean ones. In line with the dual-label approach, there is an additional target which is either a clean pitch track or a clean FFT spectrogram. Note that one significant difference between these two additional targets is that the pitch track is one dimensional whereas the FFT used creates 100 dimensions. Thus one of these additional targets has much lower dimensionality than the 31 dimensional MFB features and one has much higher dimensionality. Note that the secondary labels only assist the training of LSTMs, and ultimately the networks are used only to produce dereverberated MFB outputs.

### 2.1. Network structure

#### 2.1.1. LSTM specifics

Bidirectional LSTMs were chosen because they can take advantage of long-term both previous and future reverberant speech input to predict the current clean label. The long-term property is consistent with the signal property of reverberation.

The dimension of the hidden-hidden weights, namely the number of cells, is 256, and the number of hidden layers is 4. Both numbers were heuristically chosen, based on some preliminary experiments. The input size of the network is dictated by the dimension of MFB outputs, i.e., 31.

*2.1.2. Loss function*

Mean square error (MSE) was chosen as the loss function and the loss on both targets are weighted equally as given in Equation (1). The MSE loss function has the advantage of computational simplicity. With the introduction of secondary targets which have different dimensionality than the inputs, MSE values noticeably decrease during training, which gives researchers feedback about the network performance.

$$Loss = .5 \times Loss(Target1) + .5 \times Loss(Target2) \quad (1)$$

*2.1.3. Batch normalization*

Batch Normalization is applied right after every LSTMs layer to adjust and scale the activations.

*2.1.4. Hidden layers for secondary target*

In Figure 1, The dashed-line box labeled "Hidden Layers" illustrates that two additional linear layers precede the secondary target, when the secondary target is an FFT spectrogram. Because the input MFB features and FFT spectrograms are different in dimensionality, the extra hidden layers are helpful to make the mapping more accurate. The number of hidden layers was heuristically chosen.

## 2.2. Input and targets of networks

*2.2.1. MFB outputs as sources and primary targets*

Table 1 lists the specifics of the MFB outputs which serve as inputs and primary targets. The same frame length and frame space were used for all features used in the present work.

Table 1: *Parameters for MFB features.*

| Parameter | Value |
|---|---|
| Frame Length | 25 ms |
| Frame Space | 10 ms |
| Number of Mel Filters | 31 |
| Pre-emphasis coefficient | 0.97 |

*2.2.2. Pitch as secondary target*

Pitch tracks extracted from clean speech by the YAAPT pitch estimator [14] were used as an auxiliary set of labels. Figure 2 shows a YAPPT estimated pitch of both clean (top) and reverberant (bottom) version of an utterance. Many researches have shown that the low-frequency pitch is less susceptible to reverberation than high-frequency components. In Figure 2, although reverberation smeared the speech energy contours, the pitch track is still relatively intact and is detectable by YAAPT. We hypothesized that using pitch as a secondary target would improve reverberant to clean mappings in the low frequency range.

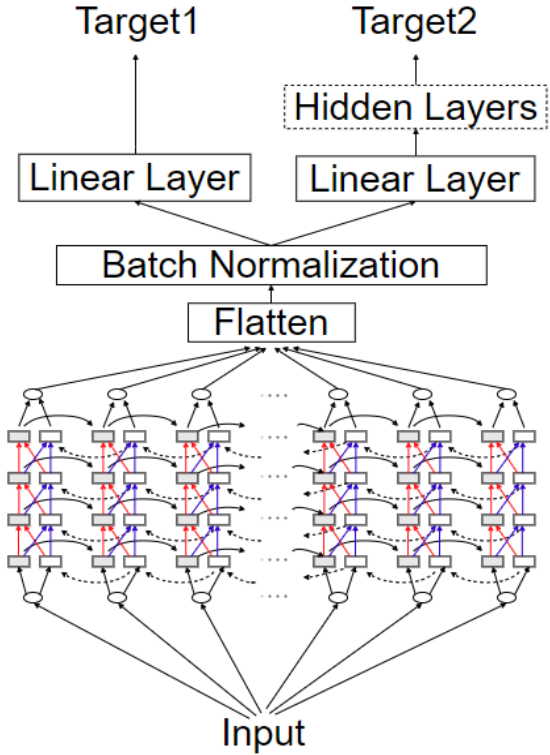

Figure 1: *LSTMs architecture.*

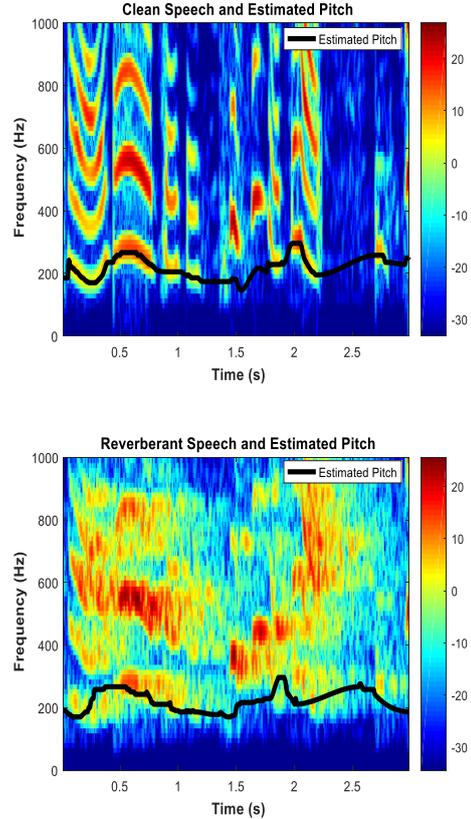

Figure 2: *YAPPT pitch tracking results*

### 2.2.3. Spectrogram as secondary target

FFT spectrograms of clean speech were also used as an auxiliary set of labels. The number of frequency bins at every frame was chosen to be 100 so the spectrogram has reasonably high resolution but does not require excessive computations for the network training.

One motivation for choosing spectrogram as secondary label is to pursue the performance gain from the cross-frequency-domain mapping [4] and also probe the reason behind the gain.

## 3. Database

Telephone speech from the Mixer 6 Database [1] was used for speaker verification tests. For this real telephone speech, a caller (channel 1) stays silent approximately half of the time and listens to the other caller (channel 2). To remove the large silent gaps in each telephone channel, voice activity detection (VAD) algorithms were employed. Then 30 seconds of continuous speech were extracted from the beginning of every VAD processed channel, which served as a 30 second sample for a speaker. In this way, 8820 equal-length sentences from 594 speakers (302 females and 292 males) were prepared.

For testing, artificial reverberation was added corresponding to the large room, far microphone (T60=0.7s) condition as per the Reverb2014 challenge data [7]. Essentially, clean sentences were convolved with the room impulse response (RIP) from [7].

## 4. Experiments

From the data described in Section 3, 6010 pairs of reverberated and clean waveforms were used for training LSTMs, where clean waveforms were responsible for MFB outputs, pitch tracks and FFT spectrograms.

After training, all reverberant waveforms to be used in the speaker verification experiment were passed through the network for processing. Speech from 100 speakers who were not present in the training data were used where 10 sentences per speaker (1000 in total) were used for enrollment and 1 sentence per speaker (100 in total) was used for evaluation.

Using the network processed sentences, speech features were computed, consisting of standard 13 Mel-frequency cepstral coefficients (MFCCs), deltas and delta-deltas. As is generally done, the first cepstral coefficient was replaced by energy. There were 39 features total.

Finally, speaker verification experiments were performed using the Alize [8] iVector system which includes a universal background model (UBM) and probabilistic linear discriminant analysis (PLDA) tools. The key parameter settings for Alize are listed in Table 2.

Table 2: *Important Alize settings.*

| Parameter | Number |
| --- | --- |
| Number of UBM mixtures | 1024 |
| iVector dimension | 200 |
| PLDA Eigenvoice dimension | 100 |
| PLDA Eigenchannel dimension | 50 |

Three cases for each type of feature/dereverberation combination were evaluated. These refer to the data for training, enrollment and testing, as listed below.
- Clean data for training, enrollment and testing (CCC).
- Clean data for training and enrollment while reverberant data for testing (CCR).
- Reverberant training, enrollment and testing data (RRR).

## 5. Results and Discussions

Figure 3 illustrates an example of input, target and LSTMs processed MFB features, using one sentence. Frame indices are converted to seconds. Using min-max normalization (2), the magnitude of all features is normalized to 0 to 1 range.

$$A_{normalized} = \frac{[A-\min(A)]}{[\max(A)-\min(A)]} \qquad (2)$$

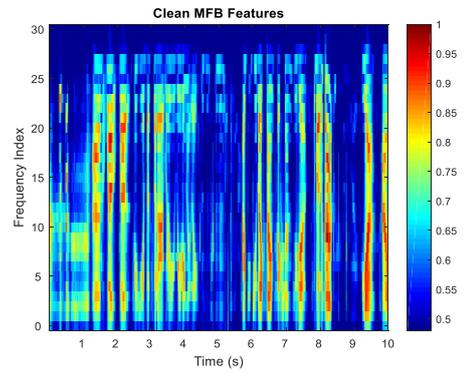

(a)

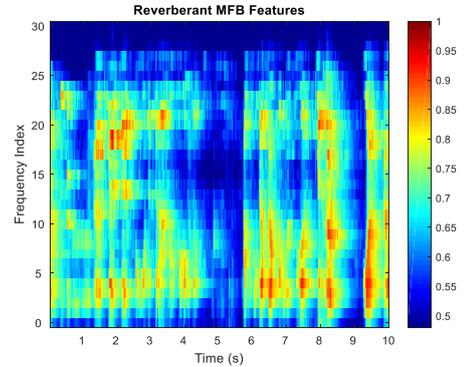

(b)

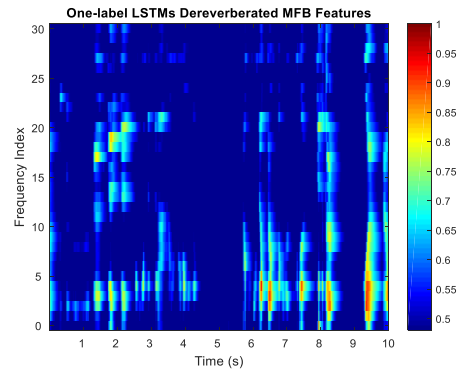

(c)

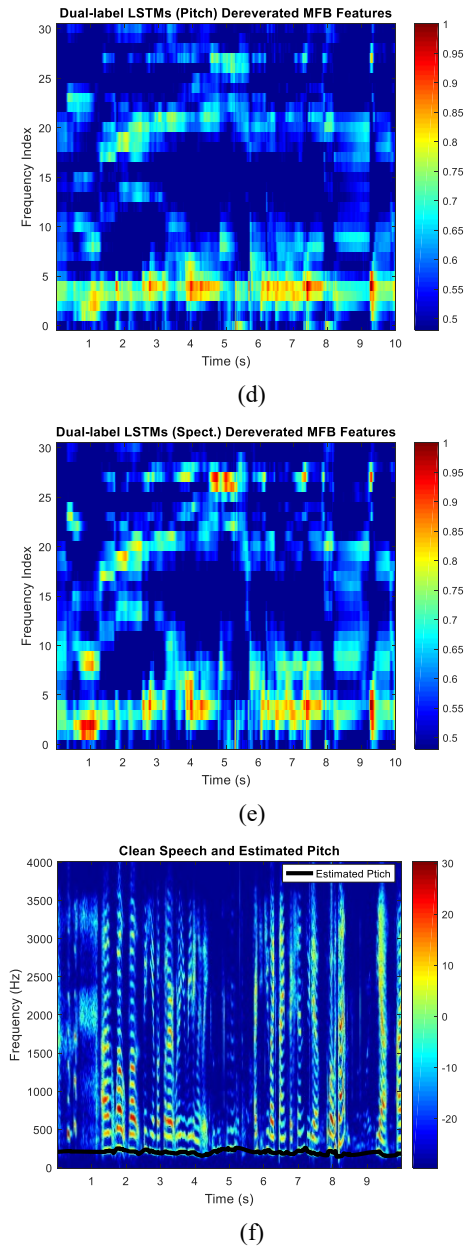

Figure 3: *MFB features and FFT spectrogram*

In Figure 3, after reverberation taints the energy distribution of MFB features to (b), one-label LSTMs can restore most of the energy contours, as in (c). However, the power in (c) is overly concentrated on the edges of each segment of speech, while many near-edge regions appear to have too low energy. The fact that high energy regions are very likely to be voiced speech important for automatic speaker verification, and that soft speech is highly distorted by reverberation and thus should deemphasized for speaker verification applications, seems to show that one-label LSTMs are not performing well for the type of preprocessing that should be used for SID tasks. Evidently, the energy-level decreases from peaks rather gradually for clean MFB features (a). In (d), dual-label LSTMs perform similarly to one-label LSTMs, while the extra pitch label causes the low-frequency energy to be more intact. Energy contours in (e) are more discontinuous than in (d) presumably because the corresponding LSTMs have different secondary labels. Possibly since a pitch track has far fewer dimensions than an FFT spectrogram, dereverberated MFB features produced by the pitch version dual-label LSTMs have smoother energy contours than those produced by the spectrogram version.

In Table 3, the second row called "One-label LSTMs" is based on using only one target which is the Mel filterbank outputs from clean utterances and the LSTMs structure is the straightforward single label version. (f) shows the estimated pitch and FFT spectrogram of the same clean sentence.

Table 3: *Dual-label LSTMs EERs (%).*

| EERs of | CCC | CCR | RRR | AVG |
|---|---|---|---|---|
| Baseline MFCCs | 2 | 13 | 6 | 7 |
| One-label LSTMs | 2 | 12.87 | **3.05** | 5.97 |
| Dual-label LSTMs pitch | 2 | **11** | 3.23 | **5.41** |
| Dual-label LSTMs spect. | 2 | 12 | **3.05** | 5.68 |

In the row called "Dual-label LSTMs pitch," the networks have two labels i) MFB outputs from clean speech ii) pitch tracks estimated using YAAPT. The "Dual-label LSTMs spect." row has the secondary targets of clean FFT spectrograms.

As shown in Table 8, the bottom two Dual-label LSTMs cases outperform their single-label counterpart by small margins, which shows the merits of the dual-label structure. Relative improvements are as follows (in EERs):

- One-label LSTMs 14.7 % reduction from baseline.
- Dual-label LSTMs pitch 24.8% reduction from baseline.
- Dual-label LSTMs spect. 18.9% reduction from baseline.
- Dual-label LSTMs pitch 4.8% reduction from Dual-label LSTM spect.

Thus the lowest EER was obtained using pitch tracks as secondary labels. Although spectrograms have higher resolution (much higher dimensionality representing complete spectrum), it is possible that the spectrogram details "misguided" the training process whose objective should be solely dereverberation rather than performing a "Mel-frequency-to-log-frequency" mapping. As a side note, dual-label LSTMs were also tested with a secondary label identical to the primary label, namely MFB features, but no benefit is discovered.

Follow on work includes testing this general approach with varying degree of reverberation.

## 6. Acknowledgements

## 7. Disclaimer

The views and conclusions contained herein are those of the authors and should not be interpreted as necessarily representing the official policies or endorsements, either expressed or implied, of the Air Force Research Laboratory or the U.S. Government.